\documentclass[pra,twocolumn,nofootinbib,floatfix,10pt]{revtex4-2}

\usepackage{amsmath}
\usepackage{amssymb}
\usepackage{wasysym}
\usepackage{graphicx}
\usepackage{color,soul}
\usepackage{physics}
\usepackage{siunitx}
\usepackage{dsfont}
\usepackage{float}
\usepackage{physics}
\usepackage{braket}
\usepackage[english]{babel}
\usepackage{blindtext}
\usepackage[english,nomargin,inline,marginclue,draft]{fixme}
\pdfpageheight\paperheight
\pdfpagewidth\paperwidth

\usepackage[colorlinks,linkcolor=blue,anchorcolor=blue,citecolor=blue,urlcolor=blue]{hyperref}

\fxusetheme{colorsig}
\FXRegisterAuthor{cg}{acg}{CG}  
\FXRegisterAuthor{th}{ath}{\color{blue}TH}  
\FXRegisterAuthor{ib}{aib}{\color{red}IB} 
\FXRegisterAuthor{sh}{ash}{\color{cyan}SH} 
\FXRegisterAuthor{db}{adb}{\color{green}DB} % now I can ..
\FXRegisterAuthor{ps}{aps}{PS}
\makeatletter
\renewcommand*\FXLayoutInline[3]{%
  {\@fxuseface{inline}\ignorespaces{\color{fx#1}[#3: #2]}}}
\makeatother

\long\def\symbolfootnote[#1]#2{\begingroup%
\def\thefootnote{\fnsymbol{footnote}}\footnotetext[#1]{#2}\endgroup}

\def\nobreakbefore{%
  \relax\ifvmode\else
    \ifhmode
      \ifdim\lastskip > 0pt\relax
        \unskip\nobreakspace
      \else % added to put a ~if no space was typed. (Unclear why it sometimes worked before )
        \nobreakspace
      \fi
    \fi
  \fi
}
\let\oldcite\cite
\renewcommand\cite{\nobreakbefore\oldcite}

\begin{document}
\title{Higher-Order Exceptional Point Sensing with Rydberg Micro-Ensemble Arrays}

\author{Jun Zhang$^{1,2,3,\textcolor{blue}{\star}}$}
\author{Ya-Jun Wang$^{1,2,3,\textcolor{blue}{\star}}$}
\author{Zi-Han Li$^{1,2,3,\textcolor{blue}{\star}}$}
\author{Jin-Hang Fang$^{4}$}
\author{Luan-Yi Wei$^{4}$}
\author{Qing Li$^{1,2}$}
\author{Guang-Can Guo$^{1,2,3}$}
\author{Bang Liu$^{1,2,4,\textcolor{blue}{\ddagger}}$}
\author{Li-Hua Zhang$^{1,2,4,\textcolor{blue}{\S}}$}
\author{Dong-Sheng Ding$^{1,2,3,4,\textcolor{blue}{\dagger}}$}
\author{Bao-Sen Shi$^{1,2,3}$}

\affiliation{$^1$Laboratory of Quantum Information, University of Science and Technology of China, Hefei, Anhui 230026, China.}
\affiliation{$^2$Anhui Province Key Laboratory of Quantum Network, University of Science and Technology of China, Hefei 230026, China.}
\affiliation{$^3$Hefei National Laboratory, University of Science and Technology of China, Hefei 230088, China.}
\affiliation{$^4$Davinci Quantum Technology Co., Ltd, Hefei, Anhui 230026, China.}

\date{\today}
\symbolfootnote[1]{J.Z, Y.J.W, Z.H.L contribute equally to this work.}
\symbolfootnote[3]{lb2016wu@ustc.edu.cn}
\symbolfootnote[4]{zlhphys@ustc.edu.cn}
\symbolfootnote[2]{dds@ustc.edu.cn}

\maketitle
\textbf{Higher-order exceptional points arise from the coalescence of multiple eigenvalues and eigenstates, leading to enhanced spectral responses to weak perturbations. However, realizing and controlling such higher-order degeneracies in interacting many-body systems remains challenging due to the complex coupling among multiple interacting eigenstates. Here, we construct spatially structured microensemble arrays in cold Rydberg gas, where long-range interaction and spatial connectivity enlarge the effective non-Hermitian dimensionality, enabling higher-order exceptional points. By varying the spatial arrangement of these sites, we realize different orders of non-Hermitian exceptional points. We further show that interactions between Rydberg excitation sites can enhance the quantum Fisher information, thereby improving sensing sensitivity. The resulting higher-order exceptional points exhibit increasingly nonlinear phase-boundary responses, which can be further tuned by varying the measurement time and atom number. By tracking shifts of these phase boundaries, we achieve a sensitivity of 0.47(10)~$\rm{nV}~cm^{–1}~Hz^{–1/2}$, surpassing the standard quantum limit. Our results establish Rydberg microensemble arrays as a platform connecting higher-order non-Hermitian criticality, many-body dynamics, and microwave sensing.}

\section*{INTRODUCTION}
Exceptional points (EPs) are non-Hermitian spectral singularities where eigenvalues and eigenstates coalesce \cite{Heiss2012,MiriAlu2019,Ozdemir2019,Ashida2020,Bergholtz2021}. EPs have been explored across photonic, microwave, optomechanical, and atomic systems \cite{Dembowski2001,Ruter2010,Xu2016,DingIon2021,Liang2023}. Owing to their anomalous spectral responses to external perturbations, EPs have further been investigated for sensing \cite{Wiersig2014,Chen2017,LauClerk2018,Langbein2018,WiersigReview2020}. Compared with lower-order EPs, higher-order EPs exhibit stronger spectral responses to weak external perturbations, which enhances the quantum Fisher information (QFI) through increased sensitivity of the quantum state~\cite{Giovannetti2006quantum,Degen2017_quantum_sensing,Frerot2018,ilias2022criticality}. Such enhanced response offers promising opportunities for precision measurement and quantum sensing~\cite{Hodaei2017,DemangeGraefe2012,WangArbitrary2019,ZhangMagnonics2019,Habler2020,xiao2026observation,mukhopadhyay2024modular}. However, realizing higher-order EPs in interacting quantum systems remains challenging due to the increased dimensionality and complex connectivity \cite{Li2023,Ding2016,Zhong2020,Mandal2021,WangSinglePhoton2023}.

Rydberg atomic systems are characterized by strong and tunable long-range interactions, which enable coupling between spatially separated excitation regions and give rise to rich many-body behavior \cite{Saffman2010quantum,Adams2020,Browaeys2020,Firstenberg2016,Defenu2023}. The spatial geometry of Rydberg arrays influences the connectivity between excitation sites, allowing control over collective dynamical evolution and many-body phases \cite{Bernien2017,Keesling2019,Ebadi2021,Scholl2021,Bluvstein2021}. In driven-dissipative Rydberg ensembles, the interplay of coherent excitation, interactions, and dissipation gives rise to phase transitions, bistability, hysteresis, and time crystal dynamics \cite{Carr2013,Marcuzzi2014,Letscher2017,Ding2020,zhang2025exceptional,liu2024bifurcation,liu2024microwave}. These nonlinear responses also provide opportunities for enhanced microwave metrology \cite{Schlossberger2024,Sedlacek2012,Jing2020,liang2026exceptional, ding2022enhanced,wang2026quantum}. Harnessing higher-order EPs to enhance microwave sensing in spatially structured Rydberg ensembles remains challenging, as the critical response depends on the interplay between spatial geometry and long-range interactions.

\begin{figure*}[t]
\centering
\includegraphics[width=1\linewidth]{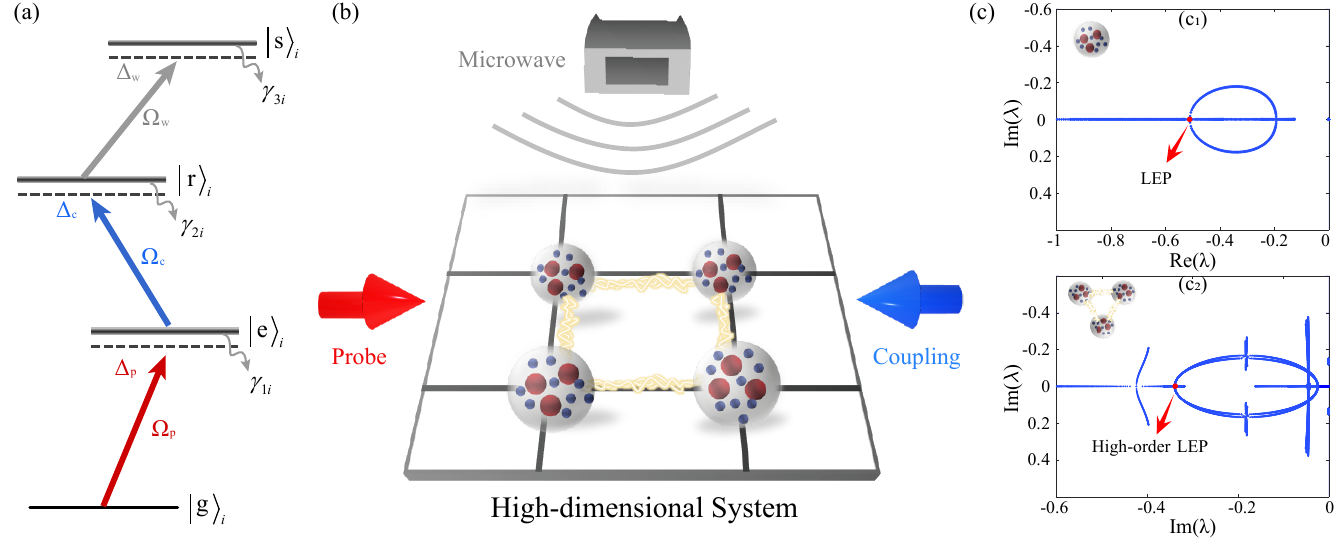}
\caption{
\textbf{Physical model and higher-order exceptional points.}
(a) Four-level scheme of $^{85}$Rb atoms at excitation site $i$, comprising $\ket{g}_i=\ket{5S_{1/2},F=3}_i$, $\ket{e}_i=\ket{5P_{3/2},F=4}_i$, $\ket{r}_i=\ket{42D_{5/2}}_i$, and $\ket{s}_i=\ket{41F_{7/2}}_i$. The probe (red), coupling (blue), and microwave (gray) fields drive the $\ket{g}_i\leftrightarrow\ket{e}_i$, $\ket{e}_i\leftrightarrow\ket{r}_i$, and $\ket{r}_i\leftrightarrow\ket{s}_i$ transitions, with Rabi frequencies $(\Omega_{\mathrm p},\Omega_{\mathrm c}, \Omega_{\mathrm w})$ and detunings $(\Delta_{\mathrm p},\Delta_{\mathrm c}, \Delta_{\mathrm w})$, respectively. (b) Experimental schematic. A spatial light modulator (SLM) shapes the probe beam into different spatial configurations, overlapping with a counterpropagating coupling beam in the cold atomic gas to form the microensemble arrays. A microwave field couples the two Rydberg states. Increasing the number of excitation sites expands the effective system dimensionality to \(4N\) for the $N$ sites microensemble arrays, forming a high-dimensional non-Hermitian system that supports higher-order EPs.
(c) Liouvillian eigenspectrum and Liouvillian exceptional points (LEPs) for ($\rm{c}_1$) a single ensemble with $(\Omega_{\rm{w}},\Omega_{\rm{c}},V_{\rm eff})=(0.3,6,0.6)\gamma$ and ($\rm{c}_2$) a three-site microensemble arrays with $(\Omega_{\rm{w}},\Omega_{\rm{c}},V_{12},V_{23},V_{13})=(0.3,6,0.1,0.1,0.1)\gamma$, both for $\Omega_{\rm{p}}\in\{0,3\gamma\}$. Higher-order LEPs emerge in the spatially coupled microensemble array.
}
\label{Fig1}
\end{figure*}

Here, we realize higher-order EPs in spatially structured cold Rydberg microensemble arrays coupled through long-range interactions, and exploit the enhanced responses for microwave sensing. By engineering the spatial connectivity of the arrays, we realize EPs of different orders, accompanied by sharper nonlinear phase boundaries at higher EP orders. We further show that interactions between excitation sites can enhance the QFI, with the enhancement depending on spatial structure and measurement time, indicating the potential for improved quantum sensing precision. The nonlinear critical response can be further tuned by varying the measurement time and atom number. We further exploit this enhanced response for microwave sensing by tracking the shift of the phase boundary, achieving a sensitivity of 0.47(10)~$\rm{nV}~cm^{–1}~Hz^{–1/2}$, surpassing the standard quantum limit (SQL). Our results demonstrate the potential of spatially structured Rydberg microensemble arrays for exploring higher-order EPs and enhancing microwave sensing.

\begin{figure*}
\centering
\includegraphics[width=1\linewidth]{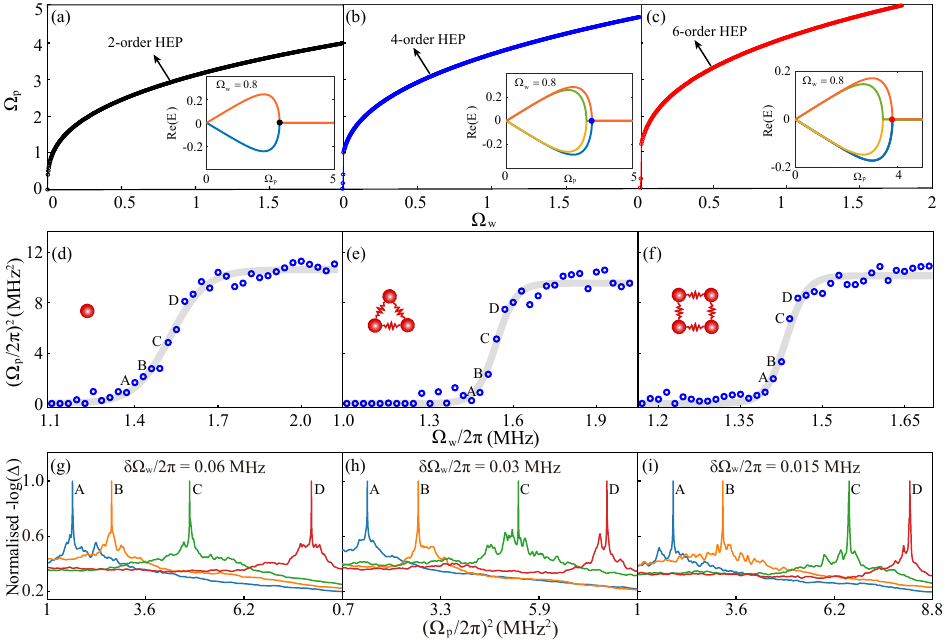}
\caption{\textbf{Nonlinear responses of higher-order exceptional points.} (a)-(c) Calculated phase boundaries for the single-site, three-site triangular, and four-site square configurations, corresponding to EP2, EP4, and EP6, respectively. The numerical parameters have been set to $\rm{\Omega_{c}} =6\gamma$ in (a),$\{\rm{\Omega_{c}},V_{12},V_{23},V_{13}\} =\{6\gamma,0.2\gamma,0.2\gamma,0.2\gamma\}$ in (b), and $\{\rm{\Omega_{c}},V_{12},V_{23},V_{34},V_{14}\} =\{12\gamma,0.4\gamma,0.4\gamma,0.4\gamma,0.4\gamma\}$ in (c). The insets show representative real parts of the Hamiltonian eigenvalues evolution at \(\Omega_{\rm{w}}=0.8\). 
(d)-(f) Measured phase boundaries for the corresponding spatial configurations. The solid lines represent sigmoid fits. The progressively steeper phase boundaries indicate an increased sensitivity to variations in the microwave field.
(g)-(i) Measured \(\log(\Delta)\) data near the critical points for the three spatial configurations in (d)-(f), with A-D corresponding to the same labeled positions, respectively.}
\label{Fig2}
\end{figure*}

\section*{RESULTS}

\subsection*{Physical model}
To investigate higher-order EP-enhanced sensing, we construct the spatially structured Rydberg microensembles arrays in cold atomic gases. Each site \(i\) comprises four energy levels, \(\ket{g}_i\), \(\ket{e}_i\), \(\ket{r}_i\), and \(\ket{s}_i\). The shaped probe field \((\Omega_{\rm{p}},\Delta_{\rm{p}})\) and a coupling field \((\Omega_{\rm{c}},\Delta_{\rm{c}})\) drive the \(\ket{g}_i\leftrightarrow\ket{e}_i\) and \(\ket{e}_i\leftrightarrow\ket{r}_i\) transitions, respectively, while a microwave field \((\Omega_{\rm{w}},\Delta_{\rm{w}})\) couples the two Rydberg states. Long-range interactions between Rydberg atoms couple spatially separated sites, enlarging the effective system dimensionality and generating a tunable non-Hermitian structure. This structure can support higher-order EPs and complex phase boundaries, providing a versatile platform for enhanced microwave sensing.

Under the mean-field approximation, the interaction between Rydberg atoms at each site is replaced by the effect of a single atom embedded in the effective field generated by the surrounding atoms. The dissipation term induced by this interaction is given by $i\Gamma_{\rm{eff}}$, where $\Gamma_{\rm{eff}}$ is proportional to the population of the $\ket{r}_i$ state~\cite{zhang2025exceptional}. In addition, long-range interactions between Rydberg atoms couple excitation sites and induce additional dissipation, whose strength depends on the spatial geometry. Accordingly, under the rotating-wave approximation, the whole non-Hermitian Hamiltonian of the microensemble arrays can be written as:
\begin{align}
H=&
\sum_{i=1}^{N} (\Omega_{\rm{p}}\hat\sigma_{eg}^{i}+\Omega_{\rm{c}}\hat\sigma_{re}^{i} +\Omega_{\rm{w}}\hat\sigma_{sr}^{i}+{\rm H.c.})+\sum_{i=1}^{N}i\Gamma_{\rm{eff}}\sigma_{rr}^i\notag\\
&-\sum_{i=1}^{N}(\sum_{j=\rm{p,c},{\rm{w}}}\Delta_j\hat{\sigma}_{ss}^{i}+\sum_{j=p,c}\Delta_j\hat{\sigma}_{rr}^{i}+\Delta_{\rm{p}}\hat{\sigma}_{ee}^{i})\notag\\
&+i\sum_{j<k}^{N}V_{jk}\,\sigma_{rr}^j\sigma_{rr}^k,
\end{align}
where $\sigma^{i}_{\alpha\beta}=\ket{\alpha}_i\bra{\beta}_i$ ($\alpha, \beta = e, g, r,s$), $N$ denotes the number of sites contained in the microensemble arrays, and $V_{jk}=V_0\,g_{jk}=V_0\, /r_{jk}^6$ characterizes the interaction-induced dissipation between microensemble sites $j$ and $k$, where $r_{jk}$ is the distance between the two sites.

For this open quantum system, the dynamics are governed by the Lindblad master equation
\begin{equation}
\dot{\rho} = -i[H, \rho] + \sum_{j} \left( L_j \rho L_j^{\dagger} - \frac{1}{2} \left\{ L_j^{\dagger} L_j, \rho \right\} \right) \equiv \mathcal{L}\rho,
\end{equation}
where $\mathcal{L}$ denotes the Liouvillian superoperator. Its spectral properties are characterized by the eigenvalue equation $\mathcal{L}\hat{\rho}_i=\lambda_i\hat{\rho}_i$, where $\lambda_i$ and $\hat{\rho}_i$ are the corresponding eigenvalues and eigenmodes, respectively. Figure~\ref{Fig1}(c) compares the Liouvillian eigenspectrum of a single-site ensemble ($\rm{c}_1$) and a spatially coupled microensemble arrays ($\rm{c}_2$). Unlike the single-site system, the spatially structured microensemble arrays support higher-order EPs, providing a new resource for enhanced microwave sensing.

To quantify the enhanced sensitivity of microwave field $\Omega_{\rm{w}}$ arising from non-Hermitian critical response in the microensemble arrays, we use the slope of parameter-response $S(\Omega_{\rm{w}},t)=\left\|P_{\perp}\ket{\partial_{\Omega_{\rm{w}}}\psi}\right\|$ with orthogonal projection $P_{\perp}=I-\ket{\psi}\bra{\psi}$~\cite{Lu2024InformationGeometry,Ren2024QuantumMetric,Zeng2025PostselectedSensing,Naikoo2026Precision}. After considering the interaction strength $V_0$ and the spatial geometry $g_{jk}$, the relative enhancement of the corresponding slope is
\begin{equation}
    \frac{S_{V_0,g_{jk}}(\Omega_{\rm{w}},t)}{S_0(\Omega_{\rm{w}},t)}\simeq 1-\frac{V_0\Gamma_{\rm eff}^2\Omega_{\rm{p}}^8}{28\,\Omega^8}\left(\frac{1}{N}\sum_{j<k}g_{jk}\right)t^3.
    \label{eq:parameter-response slope}
\end{equation}
where $\Omega=\sqrt{\Omega_{\rm{p}}^2+\Omega_{\rm{c}}^2}$ is the effective Rabi frequency, see more details in Methods section. Here, $V_0<0$, and the geometric factor $(1/N)\sum_{j<k}g_{jk}$ determines
the interaction-induced correction, leading to an
enhancement in sensitivity.

The enhanced state response can be further quantified by the QFI, which determines the quantum limit of parameter-estimation precision. Using the defination $F_Q(\Omega_{\rm{w}},t)=4|S(\Omega_{\rm{w}},t)|^2$, the QFI for different spatial configurations is given by \cite{Degen2017_quantum_sensing,Giovannetti2006quantum}
\begin{equation}
 \frac{F_Q(\Omega_{\rm{w}},t)}{N}\simeq\frac{4\,\Omega_{\rm{p}}^2}{9\,\Omega^2}t^2-\frac{V_0\Gamma_{\rm eff}^2\Omega_{\rm{p}}^{10}}{63\,\Omega^{10}}\left(\frac{1}{N}\sum_{j<k}g_{jk}\right)t^5.
\label{eq:qfi_time_EP}
\end{equation}
According to the quantum Cramér–Rao bound, $\Delta\Omega_{\rm{w}}\geq 1/\sqrt{F_Q(\Omega_{\rm{w}},t)}$, the enhancement of the QFI lowers the bound on estimation uncertainty, demonstrating the role of long-range interactions and array geometry in enhancing microwave sensing.

\begin{figure*}[t]
\centering
\includegraphics[width=1\linewidth]{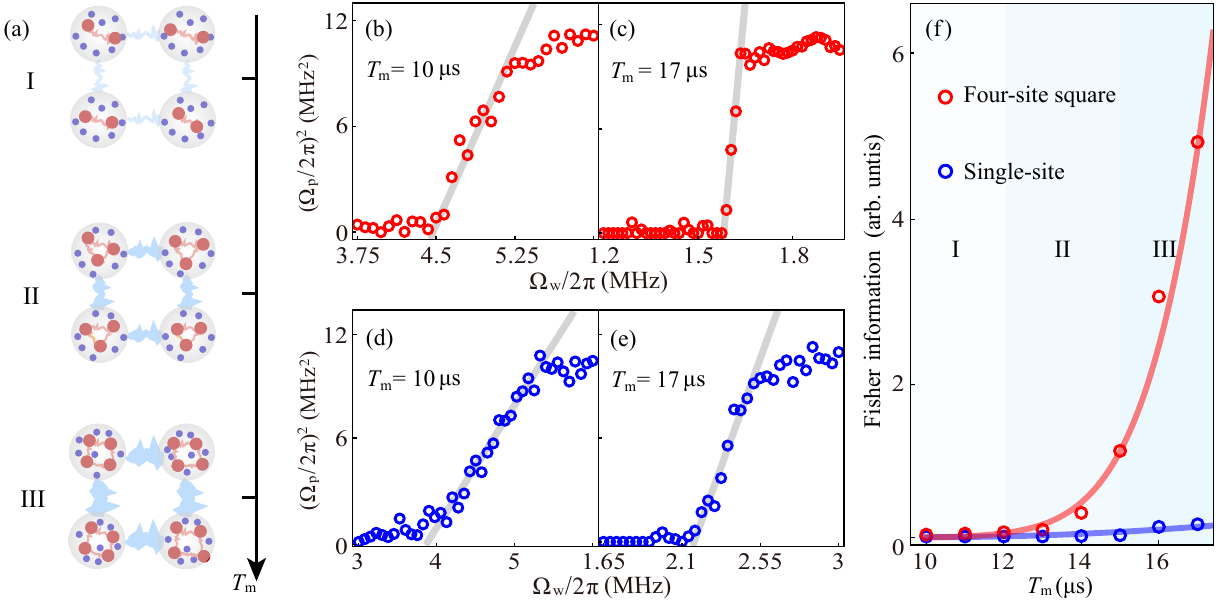}
\caption{\textbf{Dynamical evolution of Rydberg microensemble arrays.} (a) Schematic diagram of the establishment of correlation in microensemble arrays with increasing measurement time \(T_{\rm{m}}\). As \(T_{\rm{m}}\) increases, correlations are gradually develop both among atoms within each site and across the microensemble arrays. (b)-(c) Measured phase boundaries for the four-site square configuration at \(T_{\rm{m}}=10~\mu\mathrm{s}\) and \(17~\mu\mathrm{s}\), respectively. (d)-(e) Corresponding phase boundaries for the single site configuration. The gray lines indicate the slopes near the critical regime. (f) The Fisher information $F(\Omega_{\rm w},t)$
as a function of measurement time $T_{\rm m}$. The red and blue curves are fitted by function $a(T_{\rm m}-T_{\rm 0})^2+b(T_{\rm m}-T_{\rm 0})^5$ and $c(T_{\rm m}-T_{\rm 0})^2$, respectively. The enhancement of the critical response highlights the contribution of long-range interactions between Rydberg atoms at the excitation sites.}
\label{Fig3}
\end{figure*}

\subsection*{Higher-order exceptional points in Rydberg microensemble arrays}

To explore higher-order non-Hermitian criticality in long-range interacting Rydberg gases, we introduce spatial degrees of freedom by shaping the probe field into distinct excitation geometries using a spatial light modulator (SLM). Under electromagnetically induced transparency (EIT) conditions, the spatially modulated probe beam creates separated Rydberg excitation sites within the cold $^{85}\mathrm{Rb}$ ensemble. These spatial degrees of freedom increase the effective dimensionality of the system and modify its interaction connectivity, enabling the realization of higher-order EPs, see the Supplemental Materials for details of the experimental setup.

We first theoretically characterize how spatial structuring modifies the order of the EPs. For a single site, a second-order Hamiltonian EPs (HEP2) gives rise to a nonlinear phase boundary as the microwave field varies, as shown in Fig.~\ref{Fig2}(a). The insets illustrate the Hamiltonian eigenspectrum and the coalescence of eigenvalues at the corresponding EPs. Increasing the number of spatially coupled sites enlarges the effective non-Hermitian dimension, enabling the emergence of higher-order EPs. Figures~\ref{Fig2}(b) and \ref{Fig2}(c) show the nonlinear phase boundaries associated with fourth-order EPs (EP4) and sixth-order EPs (EP6) in three-site triangular and four-site square configurations, respectively. The insets in Figs.~\ref{Fig2}(a)-(c) present the real parts of the Hamiltonian eigenvalues at $\Omega_{\rm{w}}=0.8$, with black, blue, and red dots marking the EPs corresponding to EP2, EP4, and EP6, respectively. 

Figures~\ref{Fig2}(d)-(f) show the experimentally measured phase boundaries for the three spatial configurations, obtained from the flipping of hysteresis trajectories~\cite{wang2026quantum}. The phase boundaries are fitted using a sigmoid function $S(x)=B/[1+e^{-\beta(x-x_c)}]$, as indicated by the grey solid curves, where $\beta$ characterizes the steepness of the phase boundary and $x_c$ denotes the critical microwave Rabi frequency. The fitted values of $\beta$ increase from 2.3 for the single-site configuration to 4.2 for the three-site triangular configuration and 7.9 for the four-site square configuration, demonstrating a progressive enhancement of the critical response.

This enhanced response is further illustrated in Figs.~\ref{Fig2}(g)-(i), which display representative $-\log(\Delta)$ profiles near the critical region for the three spatial configurations, where $\Delta =|T_{\mathrm{up}}-T_{\mathrm{down}}|$ denotes the transmission difference between the flipping trajectories, see more detail in Supplemental Materials. The peaks identify the corresponding positions labeled A-D in Figs.~\ref{Fig2}(d)-(f), respectively. For comparable shifts in the critical regime, the required microwave Rabi frequency variation decreases from $\delta\Omega_{\rm{w}}/2\pi=0.06~\mathrm{MHz}$ for the single-site configuration to $0.03~\mathrm{MHz}$ for the three-site triangular configuration and further to $0.015~\mathrm{MHz}$ for the four-site square configuration. These results demonstrate that increasing the number of excitation sites enhances the sensitivity of the phase boundary to microwave field variations.

The enhanced response arises from the formation of higher-order EPs enabled by the spatial structure of microensemble arrays. Long-range interactions couple spatially separated excitation sites into a higher-dimensional effective system and induce additional dissipation, thereby modifying the effective non-Hermitian eigenspectrum. These allow multiple eigenvalues and eigenstates to coalesce at higher-order EPs, where small variations in the microwave field can produce pronounced spectral changes. The emergence of higher-order EPs is accompanied by experimentally observed steeper nonlinear phase boundaries and enhanced sensitivity to microwave field variations. The enhanced collective response is further reflected in Eq.~\eqref{eq:parameter-response slope} and ~\eqref{eq:qfi_time_EP}, where interactions between excitation sites modify the sensitivity of the evolved quantum state to the microwave field. These results highlight the potential of spatially structured Rydberg microensemble arrays for enhancing microwave sensing through high-order EPs.

\subsection*{Higher-order exceptional points dynamics}

The QFI increases with measurement time, highlighting the role of measurement time as a valuable resource for enhancing sensing precision\cite{pang2017optimal, manshouri2025quantum}. To investigate the dynamical evolution of of nonlinear responses near higher-order EPs, we measure the phase boundaries at different measurement times $T_{\rm m}$ for the single-site and four-site square configurations. Figure~\ref{Fig3}(a) illustrates the evolution of collective correlation in the microensemble arrays due to long-range interaction between Rydberg atoms. At short $T_{\rm m}$, correlations within each excitation site and across the microensemble arrays remain weak, which prevents the entire system from producing a strong response to $\Omega_{\rm w}$. As $T_{\rm m}$ increases, correlations among the atoms within each site are gradually established, and the correlations between sites are also gradually develop, leading to the response to $\Omega_{\rm w}$ become progressively stronger.

Figures~\ref{Fig3}(b)-(e) show the measured phase boundaries for the four-site square (red) and single-site (blue) configurations at $T_{\rm{m}}=10~\mu\mathrm{s}$ and $17~\mu\mathrm{s}$, respectively. The gray lines indicate the slopes of the phase boundaries near the critical regime. At $T_{\rm{m}}=10~\mu\mathrm{s}$, both configurations exhibit relatively small and comparable slopes. As $T_{\rm{m}}$ increases to $17~\mu\mathrm{s}$, the four-site configuration develops a markedly steeper phase boundary, whereas the single-site configuration shows only a slight increase. These results demonstrate that increasing the measurement time enhances the nonlinear response of the phase boundary, with interactions between excitation sites further amplifying this effect.

To further investigate the influence of spatial geometry on the critical response, we experimentally estimate the Fisher information $F(\Omega_{\rm w},T_{\rm m})=\kappa^2/\sigma_{\rm trans}^2$ as a function of measurement time $T_{\rm m}$, as shown in Fig.~\ref{Fig3}(f). Here, $\kappa=d\Omega_{\rm p}^2/d\Omega_{\rm w}$ represents the slope of the phase boundary in the critical region, and $\sigma_{\rm trans}^2$ denotes the variance of the transmission intensity~\cite{qsmd-yh44,ding2022enhanced}. We find that the data points are well fitted by functions $a(T_{\rm m}-T_{\rm 0})^2+b(T_{\rm m}-T_{\rm 0})^5$ and $c(T_{\rm m}-T_0)^2$ for the red and blue data, respectively, with $T_{\rm 0}=10~\mu\mathrm{s}$, $a=0.012$, $b=2.9\times10^{-4}$, and $c=0.003$. The additional fifth-order contribution in the four-site square structure indicates that long-range interactions between excitation sites further enhance the critical response. The observed time dependence of Fisher information is consistent with the scaling behavior of the QFI predicted by Eq.~\eqref{eq:qfi_time_EP}, suggesting a connection between the enhanced critical response and quantum measurement precision. At \(T_{\rm m}=17\,\mu\mathrm{s}\), the four-site square array shows an approximately 30-fold enhancement in $F(\Omega_{\rm w},T_{\rm m})$ relative to the single-site configuration. 

The distinct dynamical behaviors for nonlinearity originate from how global correlation is established in many-body system, which directly reflects how the QFI grows~\cite{Chu2023strong}. In the single-site configuration, nonlinear dynamics develop only through interactions within a single region. In contrast, the four-site square configuration involves long-range interaction between spatially separated regions, requiring time for correlation to establish across the entire arrays. At short $T_{\rm m}$, the correlation of these sites remain weak, which limits the collective nonlinear response. As $T_{\rm m}$ increases, the correlations develop both within individual sites and across the microensemble arrays, leading to a stronger nonlinear response. This mechanism underlies the progressive sharpening observed in the four-site configuration and highlights the role of global correlation in establishing higher-order EPs responses.

\begin{figure*}[t]
\centering
\includegraphics[width=1\linewidth]{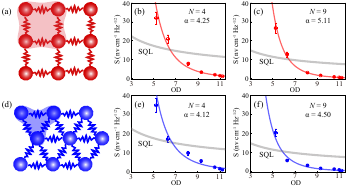}
\caption{\textbf{Sensitivity scaling with atom number across spatial configurations.}
(a) Schematic of the square spatial configuration.
(b)-(c) Measured microwave sensitivity as a function of optical depth (OD) for the square configuration with \(N=4\) and \(N=9\) excitation sites, respectively. The solid red curves are power-law fits \(S\propto \mathrm{OD}^{-\alpha}\), yielding \(\alpha=4.25\) and \(5.11\), respectively.
(d) Schematic of the triangular spatial configuration.
(e)-(f) Corresponding sensitivity measurements for the triangular configuration with \(N=4\) and \(N=9\), with fitted exponents \(\alpha=4.13\) and \(4.50\), respectively. The gray curves denote the corresponding standard quantum limit (SQL).}
\label{Fig4}
\end{figure*}

\subsection*{Higher-order exceptional points enhanced sensing}

The enhanced nonlinear response near higher-order EPs suggests potential benefits for microwave sensing. To investigate the dependence of microwave sensing sensitivity on atom number and spatial geometry, we measure the sensitivity as a function of optical depth (OD) for Rydberg microensemble arrays with different site numbers and geometries. Here, $\mathrm{OD}=N_{\rm at}\sigma_0/A$ is proportional to the atom number $N_{\rm at}$, where $\sigma_0$ is the resonant absorption cross section and $A$ is the probe beam area~\cite{zhang2012dark}. Figures~\ref{Fig4}(a) and \ref{Fig4}(d) illustrate two representative spatial configurations: a square arrangement (red) and a triangular arrangement (blue). By varying the atom number through OD and the spatial connectivity, we investigate how these factors influence the microwave response and quantum enhanced sensing performance.

Figures~\ref{Fig4}(b) and \ref{Fig4}(c) show the measured sensitivities for the square arrangement with $N=4$ and $N=9$ excitation sites, respectively. The sensitivity follows a power-law dependence on OD, $S\propto\mathrm{OD}^{-\alpha}$, with fitted exponents of $\alpha=4.25$ for $N=4$ and $\alpha=5.11$ for $N=9$, indicating a stronger OD dependence as the number of excitation sites increases. Figures~\ref{Fig4}(e) and \ref{Fig4}(f) present the corresponding results for the triangular arrangement, yielding exponents of $\alpha=4.13$ for $N=4$ and $\alpha=4.50$ for $N=9$. These distinct scaling behaviors reflect the influence of interaction strength and spatial geometry on the sensing response. In the large OD regime, the measured sensitivity surpasses the SQL. The SQL is expressed as $S_{\rm SQL}=h/(\mu_{\rm w}\sqrt{N_{\rm at}T})$\cite{bussey2022quantum}, where $h$ is Planck constant, $\mu_{\rm w}$ is the microwave transition dipole moment, and $T$ is the measurement time. Under optimized conditions, we achieve a microwave sensitivity of 0.47(10)~$\rm{nV}~cm^{–1}~Hz^{–1/2}$, corresponding to an equivalent power sensitivity of $-215.1\pm2.1~\mathrm{dBm\ Hz^{-1}}$~\cite{fancher2021rydberg}, see the Supplemental Materials for the derivation of the sensitivity.

The observed sensitivity scaling arises from the interplay of long-range interactions, higher-order EPs, and spatial connectivity between Rydberg atoms. Increasing the OD strengthens the collective interaction and enhances the nonlinear microwave response, while increasing the number of excitation sites enlarges the effective dimensionality of the non-Hermitian coupling structure, enabling higher-order EPs structures. Higher-order EPs exhibit a stronger response to weak external perturbations, resulting in a stronger dependence of the sensing sensitivity with OD. Additionally, the spatial geometry further determines the connectivity of the interaction and modifies the coupling among the collective modes, leading to different sensitivity scaling even for arrays with the same number of sites. These results demonstrate the potential of spatially structured Rydberg microensemble arrays for enhancing microwave sensitivity through long-range interactions and higher-order EPs.

\section*{Discussion}

Spatial structuring of Rydberg microensemble arrays offers further possibilities for controlling long-range interactions in complex many-body systems. Extending this approach to larger arrays with more complex spatial arrangements could enable the investigation of information spreading, spatial correlations, and higher-order interactions. Furthermore, combining programmable spatial structures with optimization algorithms could facilitate the design of interaction connectivity with specific dynamical responses. These extensions could provide a broader framework for exploring the interplay of dissipation, long-range interactions, and nonequilibrium dynamics in open quantum many-body systems.

In summary, we have proposed and experimentally demonstrated a scheme for realizing higher-order EPs using spatially structured Rydberg microensemble arrays. By controlling the number and geometry of excitation sites, we have realized higher-order EPs, accompanied by sharper nonlinear phase boundary. We futher show that this interactions between different sites can enhance the QFI, thereby improving the sensitivity of quantum sensing. By varying the measurement time, we investigated the dynamical evolution of these high-order EPs responses. We further demonstrated that microwave sensitivity depends strongly on atom number and spatial connectivity by varying the OD and spatial configuration. Under optimized conditions, we achieve a microwave sensitivity of $0.47(10)~\mathrm{nV~cm^{-1}~ Hz^{-1/2}}$, surpassing the corresponding SQL in the high OD regime. Our results demonstrate the potential of spatially structured Rydberg microensemble arrays for exploring higher-order EPs and many-body dynamics, providing a new approach to enhancing microwave sensing.

\section*{Methods}
\subsection*{Derivation of QFI for microensemble arrays}

We investigate the QFI of a spatially structured Rydberg microensemble arrays within the four-level description. Each site consists of four states, $|g\rangle$, $|e\rangle$, $|r\rangle$, and $|s\rangle$, driven by the probe, coupling, and microwave fields with Rabi frequencies $\Omega_{\rm{p}}$, $\Omega_{\rm{c}}$, and $\Omega_{\rm{w}}$, respectively. We assume resonant driving and neglect the decay of $|e\rangle$ and $|s\rangle$, while retaining the effective dissipation $\Gamma_{\rm eff}$ induced by interactions between Rydberg atoms. In the basis $\{|g\rangle,|e\rangle,|r\rangle,|s\rangle\}$, the local non-Hermitian Hamiltonian takes the form\cite{zhang2025exceptional,wang2026quantum}
\begin{equation}
h_0=
\begin{pmatrix}
0 & \Omega_{\rm{p}} & 0 & 0\\
\Omega_{\rm{p}} & 0 & \Omega_{\rm{c}} & 0\\
0 & \Omega_{\rm{c}} & -i\Gamma_{\rm eff} & \Omega_{\rm{w}}\\
0 & 0 & \Omega_{\rm{w}} & 0
\end{pmatrix}.
\label{eq:full_four_level_h}
\end{equation}
In the regime
$\Omega=\sqrt{\Omega_{\rm{p}}^2+\Omega_{\rm{c}}^2}\gg\Gamma_{\rm eff}$
with $\Omega_{\rm{c}}>\Omega_{\rm{p}}$, the corresponding
EP is approximately located at
$\Omega_{\rm w,\rm EP}=\Gamma_{\rm eff}\Omega_{\rm{p}}/4\Omega$.

For an array of $N$ identical sites, the total
non-Hermitian Hamiltonian is expressed as
\begin{equation}
H_{\rm tot}=H_0
+iV_0\sum_{j<k}g_{jk}\,
\sigma_{rr}^{(j)}\sigma_{rr}^{(k)},
\label{eq:full_four_level_array}
\end{equation}
where $H_0=\sum_{j=1}^Nh_0^j$ denotes the noninteracting Hamiltonian, while $V_0$ and
$g_{jk}$ characterize the strength and spatial dependence of the intersite interactions,
respectively. We assume that all sites are initially prepared in the Rydberg state, $|\psi_0\rangle=|r\rangle^{\otimes N}$, which can be achieved through stimulated Raman
adiabatic passage (STIRAP) prior to microwave interrogation~\cite{Vitanov2017STIRAP, Cubel2005RydbergSTIRAP}.

To evaluate the QFI associated with the microwave Rabi frequency $\Omega_{\rm{w}}$, we treat the intersite coupling $V_0$ perturbatively. Applying the Dyson expansion, the unnormalized time-evolved state can be obtained to first order as~\cite{Pan2020NonHermitianLinearResponse,Sticlet2022KuboNonHermitian,Hashemi2022LinearResponseEP}
\begin{align}
\ket{\psi(t)}=&e^{-iH_0t}\ket{\psi_0}\notag\\
&+V_0\sum_{j<k}g_{jk}\,
\int_0^t d\tau\
e^{-iH_0(t-\tau)}
\sigma_{rr}^{(k)}\sigma_{rr}^{(l)}
e^{-iH_0\tau}\ket{\psi_0}.
\end{align}
We use the slope of parameter-response $S(\Omega_{\rm{w}},t)=\left\|P_{\perp}\ket{\partial_{\Omega_{w}}\psi}\right\|$ with orthogonal projection $P_{\perp}=I-\ket{\psi}\bra{\psi}$ to quantify the sensitivity to $\Omega_{\rm{w}}$ of state~\cite{Lu2024InformationGeometry,Ren2024QuantumMetric,Zeng2025PostselectedSensing,Naikoo2026Precision}. After considering the interaction strength $V_0$ and the spatial geometry $g_{jk}$, the corresponding slope in the long time perturbative regime $t\,\Omega_{\rm w,\rm EP}\gg1$ near the EP is given by
\begin{equation}
    \frac{S_{V_0,\, g_{jk}}(\Omega_{\rm{w}},t)}{N}=\frac{\Omega_{\rm{p}} t}{3\Omega}-\frac{V_0\Gamma_{\rm eff}^2\Omega_{\rm{p}}^9}{84\,\Omega^9}\left(\frac{1}{N}\sum_{j<k}g_{jk}\right)t^4.
\end{equation}
The QFI of the corresponding process is defined as $F_Q(\Omega_{\rm{w}},t)=4|S(\Omega_{\rm{w}},t)|^2$, with the leading-order expression takes the form~\cite{Degen2017_quantum_sensing,Giovannetti2006quantum}
\begin{align}
    \frac{F_Q(\Omega_{\rm{w}},t)}{N}&=F_Q^{(0)}(\Omega_{\rm{w}},t)+F_Q^{(1)}(\Omega_{\rm{w}},t)\notag\\
    &\simeq\frac{4\,\Omega_{\rm{p}}^2}{9\,\Omega^2}t^2-\frac{V_0\Gamma_{\rm eff}^2\Omega_{\rm{p}}^{10}}{63\,\Omega^{10}}\left(\frac{1}{N}\sum_{j<k}g_{jk}\right)t^5,
\end{align}
where the first term represents the contribution from independent sensing sites, while the second captures the leading correction arising from intersite interactions. For additional dissipative loss ($V_0<0$), the interaction-induced correction is positive, indicating an enhancement of the conditional QFI. In particular, the correction is governed by the normalized geometric factor $(1/N)\sum_{j<k}g_{jk}$, demonstrating that spatial arrangement provides an additional degree of freedom for controlling the metrological response. This geometric dependence reflects the pairwise nature of the first-order interaction and highlights the role of arrays connectivity in quantum-enhanced sensing.

\subsection*{Experimental setup} 
To study higher-order EPs in spatially structured Rydberg ensembles, we prepare a cold ensemble of \(^{85}\mathrm{Rb}\) atoms in a three-dimensional magneto-optical trap (MOT) and excite the atoms to a Rydberg state using a two-photon EIT scheme. A 780-nm probe field drives the transition from the ground state $|g\rangle = |5S_{1/2},F=3\rangle$ to the intermediate excited state $|e\rangle = |5P_{3/2},F=4\rangle$, while a counter-propagating 480-nm coupling field connects the intermediate state to the Rydberg state $|r\rangle = |42D_{5/2}\rangle$. The probe beam has a waist of approximately 6.4$\mu m$ and is spatially modulated by an SLM to generate different excitation geometries. The generated spatial patterns are monitored by an intensified charge-coupled-device (ICCD) camera to verify the position and geometry of the excitation sites. The transmitted probe signal is detected by a photomultiplier tube (PMT), while a microwave field generated by an RF source and radiated through a horn antenna couples $|r\rangle$ to another Rydberg state $|s\rangle = |41F_{7/2}\rangle$ at a transition frequency of $2\pi \times 31.88~\mathrm{GHz}$, see more details in Supplemental Materials.

The hysteresis dynamics are measured by modulating the probe intensity with a triangular waveform, producing an upward scan followed by a downward scan of the probe Rabi frequency. The hysteresis process is recorded at different microwave Rabi frequencies to obtain the evolution of the flipping trajectories and extract the corresponding phase boundaries. The measurement sequence is automated to ensure consistent control of the probe scan, microwave field, and data acquisition.

\hspace*{\fill}

\section*{Acknowledgements}
We acknowledge funding from the National Key R and D Program of China (Grant No. 2022YFA1404002), the National Natural Science Foundation of China (Grant Nos. T2495253, 62435018, 125B2085).

\section*{Data Availability}
All experimental data used in this study are available from the corresponding author upon request.

\section*{Code availability}
The custom codes used to produce the results presented in this paper are available from the corresponding authors upon request.

\end{document}

\title{Supplementary Materials for ``Higher-Order Exceptional Point Sensing with Rydberg Micro-Ensemble Arrays"}

\author{Jun Zhang$^{1,2,3,\textcolor{blue}{\star}}$}
\author{Ya-Jun Wang$^{1,2,3,\textcolor{blue}{\star}}$}
\author{Zi-Han Li$^{1,2,3,\textcolor{blue}{\star}}$}
\author{Jin-Hang Fang$^{4}$}
\author{Luan-Yi Wei$^{4}$}
\author{Qing Li$^{1,2}$}
\author{Guang-Can Guo$^{1,2,3}$}
\author{Bang Liu$^{1,2,4,\textcolor{blue}{\ddagger}}$}
\author{Li-Hua Zhang$^{1,2,4,\textcolor{blue}{\S}}$}
\author{Dong-Sheng Ding$^{1,2,3,4,\textcolor{blue}{\dagger}}$}
\author{Bao-Sen Shi$^{1,2,3}$}

\affiliation{$^1$Laboratory of Quantum Information, University of Science and Technology of China, Hefei, Anhui 230026, China.}
\affiliation{$^2$Anhui Province Key Laboratory of Quantum Network, University of Science and Technology of China, Hefei 230026, China.}
\affiliation{$^3$Hefei National Laboratory, University of Science and Technology of China, Hefei 230088, China.}
\affiliation{$^4$Davinci Quantum Technology Co., Ltd, Hefei, Anhui 230026, China.}

\date{\today}
\symbolfootnote[1]{J.Z, Y.J.W, Z.H.L contribute equally to this work.}
\symbolfootnote[3]{lb2016wu@ustc.edu.cn}
\symbolfootnote[4]{zlhphys@ustc.edu.cn}
\symbolfootnote[2]{dds@ustc.edu.cn}

\maketitle

\large

The Supplemental Materials provide additional details supporting the main text, including the derivation of the quantum Fisher information (QFI); the calculation of the Liouvillian spectrum; the experimental setup; the generation of Rydberg microensemble arrays using a spatial light modulator (SLM); the estimation of microwave sensing sensitivity; and the demonstration of long-range interactions between Rydberg microensembles.

\section{Enhanced response induced by long-range Rydberg interactions}
In the Rydberg microensemble arrays, near the exceptional points (EPs), the response of the whole system to external perturbation increases dramatically. To characterize the sensitivity of such response, we perform a detailed theoretical derivation of the QFI near the EPs. In addition, we have analyzed the influence of different geometric structures of the Rydberg microensemble arrays on the measurement precision through the QFI expression.

\subsection{Derivation of quantum Fisher information for microensemble arrays}
In this subsection, we will derive the QFI of the interacting Rydberg microensemble arrays. The energy level for each atom considered here contains $|g\rangle$, $|e\rangle$, $|r\rangle$, and $|s\rangle$, resonantly driven by the probe, coupling, and microwave fields with Rabi frequencies $\Omega_{\rm p}$, $\Omega_{\rm c}$, and $\Omega_{\rm w}$, respectively. We divide the Hamiltonian $H_{\rm tot}$ for the entire system into two main parts: the single-site part and the intersite coupling part,
\begin{equation}
    H_{\rm tot}=\sum_{j=1}^N\sum_{\nu} h_{\nu}^{(j)}+V_0\sum_{j<k}^N\sum_{\nu,\lambda}g_{jk,\nu\lambda}\,\sigma_{rr,\nu}^{(j)}\sigma_{rr,\lambda}^{(k)}.
\end{equation}
Here, the subscripts \(j,k\) denote summation over different sites, with the upper limit being the total number of sites \(N\). The subscripts \(\nu,\lambda\) denote summation over the atoms within each site. The geometric factor is given by \(g_{jk,\nu\lambda} = (1/r_{jk,\nu\lambda})^6\), where \(r_{jk,\nu\lambda}\) is the spatial distance between the \(\nu\)-th atom in the \(j\)-th site and the \(\lambda\)-th atom in the \(k\)-th site. Since the sites are spatially separated from one another, and the intersite coupling $V_0$ is much weaker than the intrasite coupling between atoms, we can perform a coarse-graining procedure on the above Hamiltonian. The effective Hamiltonian for the single site can be simplified as the four-level one:~\cite{zhang2025exceptional,wang2026quantum}
\begin{equation}
h_0=
\begin{pmatrix}
0 & \Omega_{\rm p} & 0 & 0\\
\Omega_{\rm p} & 0 & \Omega_{\rm c} & 0\\
0 & \Omega_{\rm c} & -i\Gamma_{\rm eff} & \Omega_{\rm w}\\
0 & 0 & \Omega_{\rm w} & 0
\end{pmatrix},
\label{eq:full_four_level_h}
\end{equation}
where $\Gamma_{\rm eff}$ is the intrasite interaction induced dissipation~\cite{zhang2025exceptional,wang2026quantum}. In the regime $\Omega=\sqrt{\Omega_{\rm p}^2+\Omega_{\rm c}^2}\gg\Gamma_{\rm eff}$ with $\Omega_{\rm c}>\Omega_{\rm p}$, we can find an EP for above single-site Hamiltonian as
\begin{equation}
    \Omega_{w,\rm EP}=\Gamma_{\rm eff}\Omega_{\rm p}/4\Omega.
\end{equation}

Besides the effective intrasite Hamiltonian, the spatial separation between sites can be regarded as much larger than the size of each individual site, which means $g_{jk,\nu\lambda}\simeq g_{jk}$. Therefore, after coarse-graining, the inter-site interaction term can be reduced to 
\begin{equation}
    H_{\rm int}=iV_0\sum_{j<k}g_{jk}\,
\sigma_{rr}^{(j)}\sigma_{rr}^{(k)}
\end{equation}
For the whole microensemble arrays consisting of $N$ identical sites, the total non-Hermitian Hamiltonian can be divided into two independent part:
\begin{equation}
H_{\rm tot}=H_0
+iV_0\sum_{j<k}g_{jk}\,
\sigma_{rr}^{(j)}\sigma_{rr}^{(k)},
\label{eq:full_four_level_array}
\end{equation}
where $H_0=\sum_{j=1}^Nh_0^j$ denotes the non-interacting Hamiltonian, while $V_0$ and $g_{jk}$ characterize the strength and geometric dependence of the intersite interaction, respectively. We assume that all sites are initially prepared in the Rydberg state, $|\psi_0\rangle=|r\rangle^{\otimes N}$, which can be achieved through stimulated Raman adiabatic passage (STIRAP) prior to microwave interrogation~\cite{Vitanov2017STIRAP, Cubel2005RydbergSTIRAP}.

To evaluate the QFI associated with the microwave Rabi frequency $\Omega_{\rm w}$, we treat the intersite coupling $V_0$ perturbatively. Applying the Dyson expansion, the unnormalized time-evolved state can be obtained to first order as~\cite{pan2020non,Sticlet2022KuboNonHermitian,Hashemi2022LinearResponseEP}
\begin{align}
\ket{\psi(t)}=&e^{-iH_0t}\ket{\psi_0}+V_0\sum_{j<k}g_{jk}\,
\int_0^t d\tau\
e^{-iH_0(t-\tau)}
\sigma_{rr}^{(k)}\sigma_{rr}^{(l)}
e^{-iH_0\tau}\ket{\psi_0}+O(V_0^2).
\end{align}
We use the slope of parameter-response $S(\Omega_{\rm w},t)=\left\|P_{\perp}\ket{\partial_{\Omega_{w}}\psi}\right\|$ with orthogonal projection $P_{\perp}=I-\ket{\psi}\bra{\psi}$ to quantify the sensitivity to $\Omega_{\rm w}$ of state~\cite{Lu2024InformationGeometry,Ren2024QuantumMetric,Zeng2025PostselectedSensing,Naikoo2026Precision}. After considering the interaction strength $V_0$ and the spatial geometry $g_{jk}$, the corresponding slope in the long-time perturbative regime $t\,\Omega_{w,\rm EP}\gg1$ near the EP is given by
\begin{equation}
    \frac{S_{V_0,\, g_{jk}}(\Omega_{\rm w},t)}{N}=\frac{\Omega_{\rm p} t}{3\Omega}-\frac{V_0\Gamma_{\rm eff}^2\Omega_{\rm p}^9}{84\,\Omega^9}\left(\frac{1}{N}\sum_{j<k}g_{jk}\right)t^4+O(V_0^2).
\end{equation}
It can be found that the response to the microwave $\Omega_{\rm w}$ for quantum state is enhanced by introducing the interaction induced dissipation between different sites. Furthermore, the QFI of the corresponding process is defined as $F_Q(\Omega_{\rm w},t)=4|S(\Omega_{\rm w},t)|^2$, with the leading-order expression takes the form
\begin{align}
    \frac{F_Q(\Omega_{\rm w},t)}{N}&=F_Q^{(0)}(\Omega_{\rm w},t)+F_Q^{(1)}(\Omega_{\rm w},t)+O(V_0^2)\notag\\
    &\simeq\frac{4\,\Omega_{\rm p}^2}{9\,\Omega^2}t^2-\frac{V_0\Gamma_{\rm eff}^2\Omega_{\rm p}^{10}}{63\,\Omega^{10}}\left(\frac{1}{N}\sum_{j<k}g_{jk}\right)t^5+O(V_0^2),
    \label{eq_SM:qfi}
\end{align}
where the first term represents the contribution from independent sensing sites, while the second captures the leading correction arising from intersite interactions and the geometric structure of the whole arrays. For additional dissipative loss ($V_0<0$), the interaction-induced correction is positive, indicating an enhancement of the conditional QFI. In particular, the correction is governed by the normalized geometric factor $(1/N)\sum_{j<k}g_{jk}$, demonstrating that spatial arrangement provides an additional degree of freedom for controlling the metrological response. This geometric dependence reflects the pairwise nature of the first-order interaction and highlights the role of arrays connectivity in quantum-enhanced sensing. Finally, according to the quantum Cramer-Rao inequality
\begin{equation}
    \Delta\Omega_{\rm w}\geq\frac{1}{\sqrt{F_Q(\Omega_{\rm w},t)}},
\end{equation}
the estimation precision for microwave amplitude $\Omega_{\rm w}$ can be increased by introducing the interaction and geometric structure. The microensemble array with interactions can achieve a measurement precision for $\Omega_{\rm w}$ scaling as $1/t^{5/2}$, which has overcome the parameter sensing limit by the non-interacting case $1/t$. 

\subsection{Geometric dependence of quantum Fisher information}
\begin{figure}[t]
    \centering
    \includegraphics[width=\columnwidth]{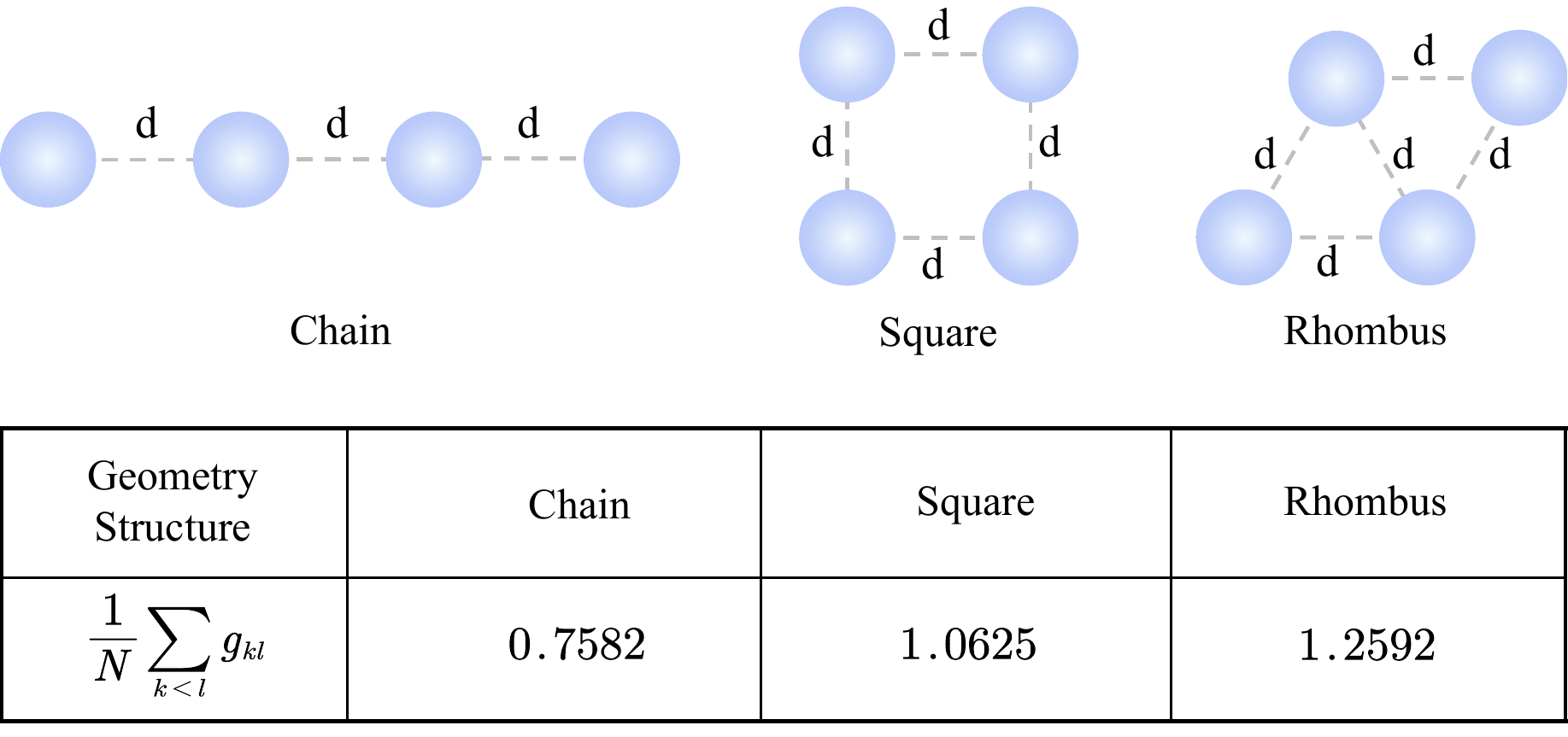}
    \caption{\textbf{Geometry structure-dependent QFI.} Spatial configurations of four sensing sites arranged in a linear chain, a square, and a rhombus, with minimum intersite separation d. The corresponding normalized geometric factors $G=N^{-1}\sum_{k<l}g_{kl}$ are $0.7582$, $1.0625$, and $1.2593$, respectively.
    }
    \label{S1}
\end{figure}

We further investigate the influence of different geometric structure on the metrological performance of interacting non-Hermitian sites. As illustrated in Fig.~\ref{S1}, we consider three representative configurations of 4 identical sites: linear chain, square and rhombus. The minimum intersite separation is fixed at $d$, and the dimensionless distance-dependent geometric factor is noted by $g_{kl}=(d/r_{kl})^6$, where $r_{kl}$ denotes the distance between sites $k$ and $l$, and scaling represents the Van der Waals interaction in Rydberg atoms. The reason for introducing the dimensionless geometric factor $g_{jk}$ here is that it allows us to conveniently compare the effects of different configurations on the response performance.

According to the non-Hermitian linear-response theory derived above, the leading interaction correction to the QFI depends on the spatial configuration through the normalized geometric factor
\begin{equation}
G
=
\frac{1}{N}\sum_{k<l}g_{kl}
=
\frac{1}{N}\sum_{k<l}
\left(\frac{d}{r_{kl}}\right)^6.
\label{eq:geometry_factor}
\end{equation}
For the chain configuration, three pairs are separated by $\rm d$, two by $2d$, and one by $3d$, yielding $G_{\mathrm{chain}}=(3+2/64+1/729)/4\simeq0.7582$. For the square configuration, four pairs are separated by $a$ and two by $\sqrt{2}d$, giving $G_{\mathrm{square}}=17/16=1.0625$. For the rhombic configuration with an interior angle of $60^\circ$, five pairs are separated by $ d$ and one by $\sqrt{3} d$, resulting in $G_{\mathrm{rhombus}}=(5+1/27)/4\simeq1.2593$. Consequently, the three configurations exhibit the ordering $F_{Q,\mathrm{rhombus}}>F_{Q,\mathrm{square}}>F_{Q,\mathrm{chain}}$ within this perturbative time window. The rhombic configuration produces approximately $1.66$ times the first-order QFI correction of the linear chain, demonstrating that spatial rearrangement alone can substantially modify the parameter sensitivity without changing the number of sensing sites.

These results establish a direct connection between array geometry and quantum metrological response, with compact configurations exhibiting stronger cumulative interactions and larger first-order QFI corrections in the considered regime.

\begin{figure}[t]
    \centering
    \includegraphics[width=\columnwidth]{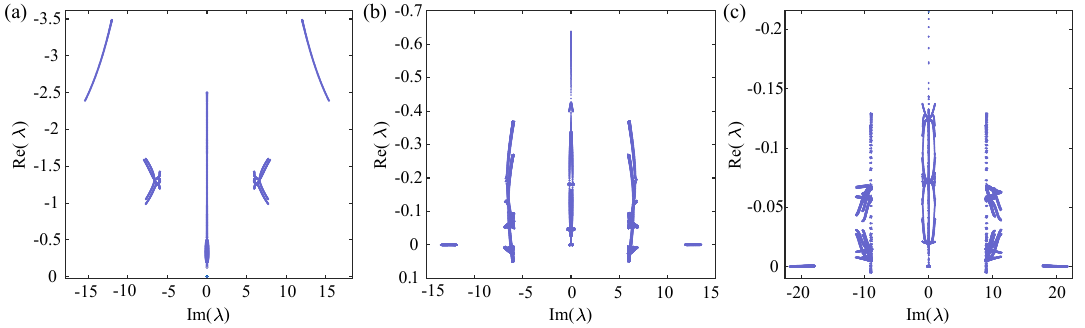}
    \caption{\textbf{The Liouvillian eigenspectrum complex planes.} The Liouvillian eigenspectrum complex planes of the system are plotted as a function of $\rm{\rm{\Omega_{p}}}$, with the ensembles having a single site shown in (a), those having three sites shown in (b), and those having four sites shown in (c). The parameters of this system are set as $\{\rm{\Omega_{c}},\rm{\Omega_{w}},V_{\rm eff}\} =\{6,0.3,0.6\}\gamma$ in (a),  $\{\rm{\Omega_{c}},\rm{\Omega_{w}},V_{12},V_{23},V_{13}\}=\{6,0.3,0.1,0.1,0.1\}\gamma$ in (b), and $\{\rm{\Omega_{c}},\rm{\Omega_{w}},V_{12},V_{23},V_{34},V_{14}\}=\{9,0.9,0.1,0.1,0.1,0.1\}\gamma$ in (c).}
    \label{S2}
\end{figure}

\section{Liouvillian spectrum for different arrays}
To further investigate the influence of spatial geometry on the non-Hermitian spectral properties of the Rydberg microensemble arrays, we calculate the Liouvillian eigenspectrum for three configurations. The dynamics of the open quantum system are governed by the Lindblad master equation: 
\begin{equation}
\frac{d\rho}{dt}=-i[H,\rho]+\sum_j\left(
L_j\rho L_j^\dagger
-\frac{1}{2}\{L_j^\dagger L_j,\rho\}
\right),
\end{equation}
where $H$ describes the coherent driving and intersite interactions, and ($L_j$) denotes the corresponding dissipation channels. The Liouvillian eigenspectrum is obtained by solving $\mathcal{L}\hat{\rho}_i = \lambda_i \hat{\rho}_i$, where $\lambda_i$ and $\hat{\rho}_i$ denote the Liouvillian eigenvalues and eigenstates, respectively. It is worth emphasizing that EPs of the Hamiltonian and those of the Liouvillian describe distinct aspects of non-Hermitian dynamics. The imaginary parts of the complex eigenenergies of an effective non-Hermitian Hamiltonian are generally associated with the decay rates of the corresponding states, whereas the Liouvillian eigenvalues characterize the dynamical evolution of the density matrix. Specifically, the real and imaginary parts of a Liouvillian eigenvalue determine, respectively, the decay rate and oscillation frequency of the associated dynamical mode. Therefore, the Liouvillian spectrum provides information about both the relaxation dynamics and coherent oscillations of the open quantum system, which cannot be directly inferred from the Hamiltonian spectrum. Moreover, a Liouvillian EP is characterized by the coalescence of both eigenvalues and their corresponding eigenmodes, and should therefore be distinguished from a Hamiltonian EP. 

Figure~\ref{S2} shows the Liouvillian eigenspectrum complex planes of the  microensemble arrays with a single site, three sites, and four sites. The real parts of these eigenvalues reflect the decay of the eigenmodes of the microensemble, while the imaginary parts reflect the oscillation frequencies of the eigenmodes. We are more concerned with the low-frequency part where the imaginary parts of the Liouvillian eigenvalues are close to zero. The Liouvillian exceptional points (LEPs) arising from the coalescence of eigenvalues and eigenstates appear in this low-frequency part. For microensembles with three or four sites, higher-order LEPs appear near the low-frequency region due to the increase in the system's degrees of freedom. Moreover, it reveals a remarkable transition in the characteristics of spontaneous symmetry breaking as the system traverses a gap-closing point.  

\begin{figure}[t]
    \centering
    \includegraphics[width=\columnwidth]{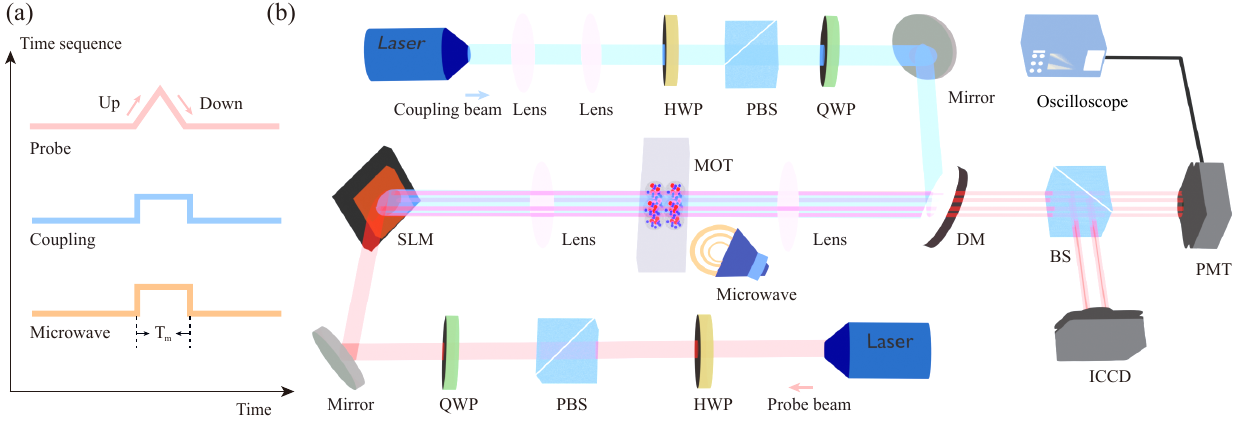}
    \caption{\textbf{Experimental setup and time sequence.} (a) The time sequence of the experiment, in which the probe beam is modulated with a triangular wave to generate hysteresis trajectories. (b) The diagram of the experimental setup. The probe beam is incident opposite to the coupling beam and focused in cold $^{85}$Rb atoms. A horn radiates microwave electric field on atoms. The following optical components are briefly described in the figure: spatial light modulator (SLM), half-wave plate (HWP), quarter-wave plate (QWP), polarizing beam splitter (PBS), dichroic mirror (DM), and photomultiplier tube (PMT), intensified charge-coupled device (ICCD).}
    \label{S3}
\end{figure}

\section{Experimental setup of cold Rydberg microensemble arrays}
To study high-order EPs enhanced sensing, we prepare a cold ensemble of $^{85}$Rb atoms trapped in a three-dimensional magneto-optic trap, as shown in Fig.~\ref{S3}. The atoms are optically pumped into the ground state $\ket{g}=\ket{5S_{1/2},F=3}$. Rydberg excitation is achieved through a two-photon transition, in which the probe laser couples $\ket{g}$ to the intermediate state $\ket{e}=\ket{5P_{3/2},F=4}$, while the counter-propagating coupling laser drives the transition from $\ket{e}$ to $\ket{r}=\ket{42D_{5/2}}$. A microwave field emitted from a horn antenna couples $\ket{r}$ to $\ket{s}=\ket{41F_{7/2}}$, with a transition frequency of $2\pi\times$31.88 GHz. In the experiment, the detuning values for the coupling, probe, and microwave field were set to zero, which is consistent with the conditions used in the theoretical calculations.

The probe beam is spatially modulated by a SLM and focused into the cold atomic ensemble to generate Rydberg microensemble arrays with different site numbers and geometries. The spatial intensity distribution of the probe beam is monitored using an intensified charge-coupled device (ICCD) camera, while the transmitted probe light is detected by a photomultiplier tube (PMT) to obtain the electromagnetically induced transparency (EIT) transmission signal. The generation of the spatial excitation patterns is described in the following section.

To obtain the hysteresis trajectories, the probe intensity is periodically varied using an acousto-optic modulator (AOM) driven by a function generator (RIGOL DG4102). The applied triangular waveform is given by $V(\tau)=6[1/2-|\tau/T_{\rm{m}}-1/2|]$, where $T_{\rm{m}}$ is the measurement time of one scanning cycle and $0\leq\tau\leq T_{\rm{m}}$. As illustrated in Fig.~\ref{S3}(a), this modulation produces successive upward and downward scans of $(\Omega_{\mathrm p}/2\pi)^2$. During the scans, the microwave field is applied using a computer-controlled RF source, while the transmitted probe signal is automatically recorded and saved by an oscilloscope.

\section{Generation of Rydberg microensemble arrays by SLM}
\begin{figure}[t]
    \centering
    \includegraphics[width=\columnwidth]{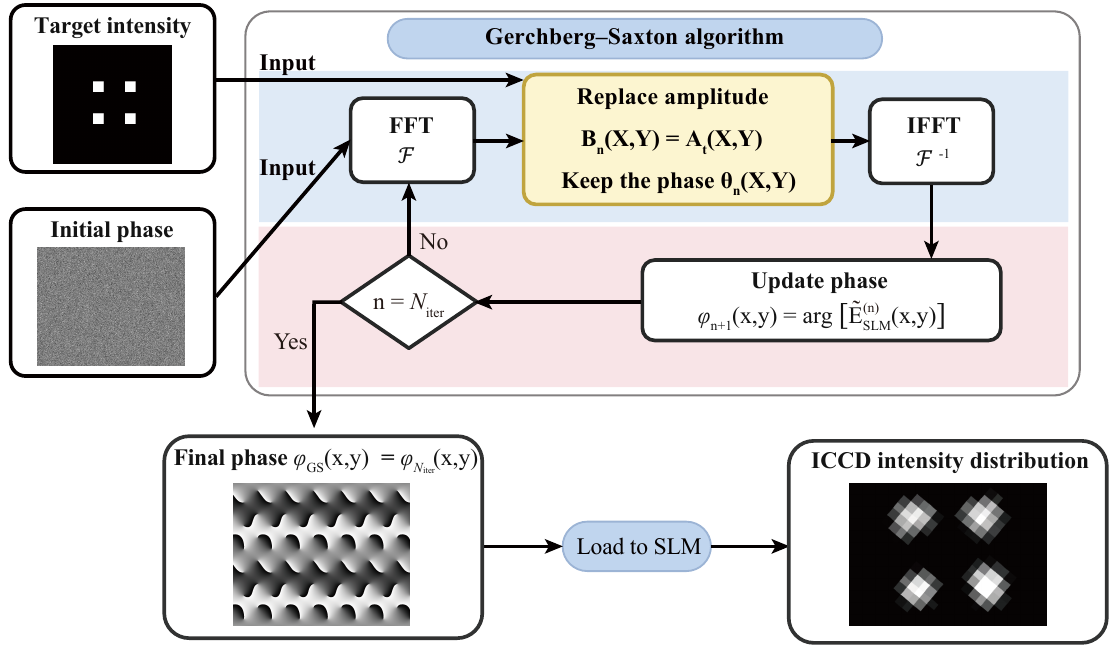}
    \caption{
    \textbf{Gerchberg-Saxton algorithm for generating the SLM phase hologram.} Starting from a random initial phase distribution, the optical field is propagated to the target plane using a fast Fourier transform (FFT). The calculated amplitude is replaced by the prescribed target amplitude while retaining the phase. An inverse fast Fourier transform (IFFT) is then applied to obtain the updated phase distribution for the next iteration. After a fixed number of iterations $N_{iter}$, the final phase hologram is loaded onto the SLM, and the resulting intensity distribution is monitored by the ICCD camera.
}
    \label{S4}
\end{figure}

The spatial configurations of the Rydberg microensemble arrays are controlled by shaping the probe laser beam using a SLM. The SLM modulates the spatial phase of the incident probe beam, and the modulated beam is focused into the atomic ensemble. The focusing lens performs a Fourier transform of the modulated optical field, converting the imposed phase distribution into the desired intensity pattern within the atomic ensemble. The intensity distribution of probe beam is imaged onto an ICCD camera.

To generate probe-beam arrays with different spatial geometries, we first define the desired arrangement of excitation sites with a tunable spacing between adjacent sites. For the multi-site configurations, the target intensity distribution is defined as
\begin{equation}
I_{\mathrm{t}}(X,Y)
=
\sum_{j=1}^{N} I_j
\delta(X-X_j,Y-Y_j),
\label{eq:GS_target}
\end{equation}
where $N$ is the number of sites, $(X_j,Y_j)$ denotes the position of the $j$-th site, $I_j$ is the corresponding intensity weight. The corresponding target amplitude distribution is denoted by $A_{\mathrm{t}}(X,Y)
=\sqrt{I_{\mathrm{t}}(X,Y)}$.

The Gerchberg–Saxton (GS) algorithm calculates the phase hologram through iterative Fourier transformations between the SLM plane and the target plane, and the framework of the GS algorithm is shown in Fig.~\ref{S4}. Starting from a randomly initialized phase distribution $\phi_0(x,y)$, a two-dimensional fast Fourier transform (FFT) is performed to obtain the optical field in the target Fourier plane,
\begin{equation}
\begin{aligned}
E^{(n)}(X,Y)
&=\mathcal{F}\left[
e^{i\phi_n(x,y)}
\right]
=B_n(X,Y)e^{i\theta_n(X,Y)},
\end{aligned}
\label{eq:GS1}
\end{equation}
where $\mathcal{F}$ denotes the Fourier transform, and $B_n(X,Y)$ and $\theta_n(X,Y)$ represent the calculated amplitude and phase in the target plane, respectively. Here, $\phi_n(x,y)$ is the phase distribution at the $n$-th iteration, with a normalized unit amplitude assumed in the SLM plane.

The calculated amplitude $B_n(X,Y)$ is then replaced by the target amplitude $A_{\mathrm{t}}(X,Y)$, while the phase $\theta_n(X,Y)$ is retained,
\begin{equation}
\widetilde{E}^{(n)}(X,Y)
=
A_{\mathrm{t}}(X,Y)e^{i\theta_n(X,Y)}.
\label{eq:GS2}
\end{equation}
The modified optical field is subsequently transformed back to the SLM plane using an inverse fast Fourier transform (IFFT),
\begin{equation}
\widetilde{E}_{\mathrm{SLM}}^{(n)}(x,y)
=
\mathcal{F}^{-1}\left[
\widetilde{E}^{(n)}(X,Y)
\right].
\label{eq:GS3}
\end{equation}
The phase of the resulting field is extracted and used for the next iteration,
\begin{equation}
\varphi_{n+1}(x,y)
=
\arg\left[
\widetilde{E}_{\mathrm{SLM}}^{(n)}(x,y)
\right].
\label{eq:GS4}
\end{equation}

The FFT and IFFT operations are repeated for a predefined number of iterations $N_{\mathrm{iter}}$. The final phase hologram is obtained as
\begin{equation}
\varphi_{\mathrm{GS}}(x,y)
=
\varphi_{N_{\mathrm{iter}}}(x,y).
\label{eq:GS5}
\end{equation}
The corresponding intensity distribution in the target plane is then numerically reconstructed as
\begin{equation}
I_{\mathrm{out}}(X,Y)
=
\left|
\mathcal{F}\left[
e^{i\varphi_{\mathrm{GS}}(x,y)}
\right]
\right|^2.
\label{eq:GS6}
\end{equation}
The calculated phase hologram is subsequently loaded onto the SLM to generate the desired probe-beam intensity pattern at the atomic ensemble. The resulting intensity distribution is imaged onto the ICCD camera for monitoring.

\section{Sensing with flipping trajectory}
\begin{figure}[t]
    \centering
    \includegraphics[width=\columnwidth]{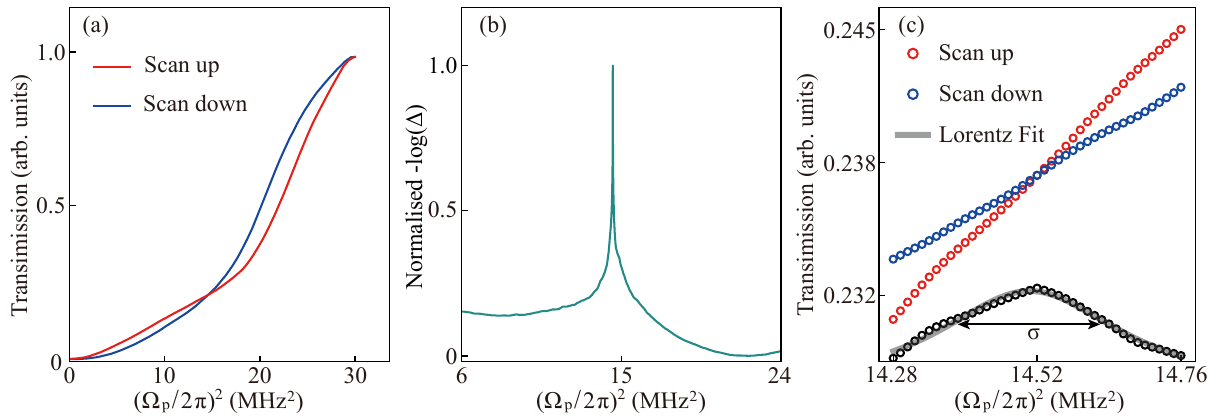}
    \caption{
    \textbf{Data processing and sensitivity estimation based on flipping trajectories.} (a) Measured probe transmission during the up (red) and down (blue) scans, showing the intersection of the hysteresis trajectories. (b) Normalized logarithmic signal obtained from the transmission difference between the scanning process. The sharp peak identifies the intersection position. (c) Transmission signals near the intersection point. The black circles represent the transmission difference, and the gray curve shows a Lorentzian fit. Here, $\sigma$ denotes the uncertainty in the intersection position.
    }
    \label{S5}
\end{figure}

The hysteresis trajectories are obtained by scanning the probe intensity $\Omega_{\rm p}^2(t)$ in the upward and downward directions. Under a microwave field, the competition between transmission and absorption leads to the flipping hysteresis trajectories, resulting an intersection between the two scanning branches, as shown in Fig.~\ref{S5}(a). Additionally, varying the microwave field shifts the intersection position, providing a measurable response for microwave field sensing~\cite{wang2026quantum}.

To identify the intersection position, we first calculate the transmission difference between the two scanning trajectories $\Delta = \left|T_{\mathrm{up}}-T_{\mathrm{down}}\right|$. A logarithmic transformation is then applied to the difference signal $ -\ln\Delta+\epsilon$, where $\epsilon$ is a small positive parameter introduced to avoid the logarithmic singularity at $\Delta = 0$. The resulting signal is normalized for visualization. As shown in Fig.~\ref{S5}(b), the intersection of the two scanning trajectories produces a sharp peak in the logarithmic spectrum, allowing the intersection position to be identified.

To evaluate the microwave sensing performance, we analyze the up and down transmission signals near the intersection, as shown in Fig.~\ref{S5}(c). The transmission difference $\Delta$ is fitted with a Lorentzian function, and the half-width at half-maximum (HWHM) of the fitted result is used to define the uncertainty $\sigma$ in the intersection position. This uncertainty is primarily attributed to photon shot noise in the measured transmission signals. 

The intersection position shifts with the applied microwave field. For a microwave field variation $\Delta E_{\mathrm w}$, the corresponding shift in $(\Omega_{\mathrm p}/2\pi)^2$ is denoted by $\delta$. The average positional uncertainty is defined as $\Sigma=\sum_{i=1}^{M}\sigma_i / M$, where $\sigma_i$ is the uncertainty associated with the $i$-th intersection point and $M$ is the total number of intersection points near the critical regime. The microwave electric-field sensitivity is then estimated as
\begin{equation}
\mathcal S=
\frac{\Sigma}{|\delta|}
\Delta E_{\mathrm w}\sqrt{T},
\label{eq:SM_sensitivity}
\end{equation}
where $T$ is the effective measurement time. The uncertainty in the estimated sensitivity arises from variations in $\sigma_i$ among these intersection points. The corresponding power sensitivity is calculated as $P_{\mathrm{sens}}= 10\log_{10}(\varepsilon_0 c\mathcal S^2 A/10^{-3})$ ~\cite{fancher2021Rydberg}, where $\varepsilon_0$ is the vacuum permittivity, $c$ is the speed of light, and $A$ is the effective detection area, which scales with the number of excitation sites.

\section{Demonstration of long-range interactions between Rydberg microensemble arrays}

\begin{figure}[t]
    \centering
    \includegraphics[width=\columnwidth]{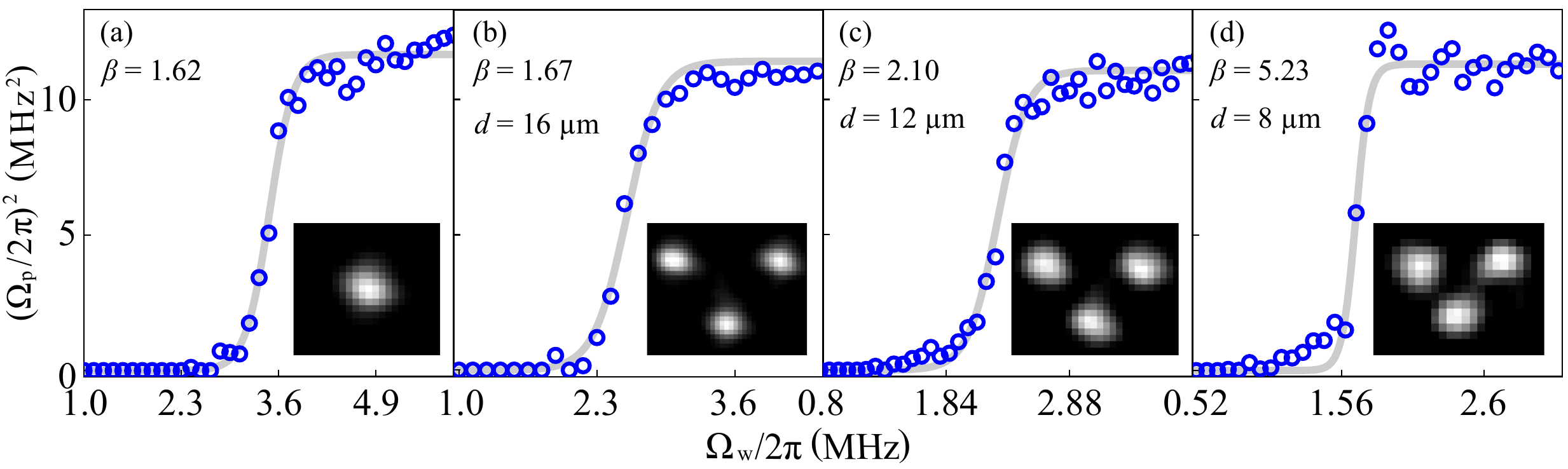}
    \caption{
    \textbf{Influence of intersite interactions on the phase boundary.} (a) Measured phase boundary for a single Rydberg microensemble. (b-d) Measured phase boundaries for three-site triangular configurations with intersite separations of $d=16$, $12$, and $8~\rm{\mu}$m, respectively. Blue open circles represent the experimental data, and gray solid curves are sigmoid fits. The corresponding steepness of the phase boundary $\beta$ are indicated in each panel. Insets show the spatial intensity distributions of the SLM-generated excitation patterns.
}
    \label{S6}
\end{figure}

To investigate long-range interactions between spatially separated Rydberg microensembles, we use the SLM to generate three-site triangular configurations with tunable intersite separation d. By varying the distance between excitation sites, we control the strength of intersite Rydberg interactions while maintaining the same number of excitation sites and spatial geometry. Figure~\ref{S6} shows the phase boundaries extracted from the hysteresis trajectories for different spatial configurations, allowing us to examine how intersite interactions influence the nonlinear response near the critical region. The insets show the spatial intensity distributions of the excitation beams shaped by the SLM. To quantitatively characterize the phase boundary, we fit the experimental data using a sigmoid function $S(x)=B/[1+e^{-\beta(x-x_c)}]$, as indicated by the grey solid curves, where $\beta$ characterizes the steepness of the phase boundary and $x_c$ denotes the critical microwave Rabi frequency.

For comparison, we first measure the phase boundary of a single Rydberg microensemble, yielding a steepness parameter of $\beta=1.62$, as shown in Fig.~\ref{S6}(a). At a large separation of $ d=16~\rm{\mu}$m, the three-site configuration exhibits a similar steepness parameter of $\beta=1.67$ [Fig.~\ref{S6}(b)], suggesting that interactions between the spatially separated excitation sites are weak. In this regime, the phase-boundary steepness of the three-site configuration remains comparable to that of a single microensemble. As the separation decreases to $d=12~\rm{\mu}$m, the phase boundary becomes steeper, with the steepness parameter increasing to $\beta=2.10$ [Fig.~\ref{S6}(c)]. A further reduction to $d=8~\rm{\mu}$m leads to a more pronounced steepening of the phase boundary [Fig.~\ref{S6}(d)]. 

These results demonstrate that reducing the intersite separation strengthens the long-range interactions between Rydberg atoms, resulting in a steeper phase boundary near the critical region. This behavior is consistent with our theoretical model, in which intersite interactions induce dissipation and enhance the QFI, highlighting the important role of long-range interactions in enhancing sensing performance.